\documentclass[letter,
               keeplastbox,   
               ]{jacow}
\usepackage{pdfpages,multirow,ragged2e} %
\makeatletter%
	\ifboolexpr{bool{xetex}}
	 {\renewcommand{\Gin@extensions}{.pdf,%
	                    .png,.jpg,.bmp,.pict,.tif,.psd,.mac,.sga,.tga,.gif,%
	                    .eps,.ps,%
	                    }}{}
\makeatother

\ifboolexpr{bool{xetex} or bool{luatex}} 
 {}                                      
 {\usepackage[utf8]{inputenc}}           

\usepackage[USenglish]{babel}

\ifboolexpr{bool{jacowbiblatex}}%
 {%
  \addbibresource{jacow-test.bib}
  \addbibresource{biblatex-examples.bib}
 }{}
\begin{document}

\title{Novel materials for next-generation\\ accelerator target facilities\thanks{This work was produced by Fermi Research Alliance, LLC under Contract No. DE-AC02-07CH11359 with the U.S. Department of Energy, Office of Science, Office of High Energy Physics. Publisher acknowledges the U.S. Government license to provide public access under the DOE Public Access Plan}}

\author{K. Ammigan\thanks{ammikav@fnal.gov}, G. Arora, S. Bidhar, A. Burleigh, F. Pellemoine\\ Fermi National Accelerator Laboratory, Batavia, IL, USA \\
		A. Couet, N. Crnkovich, I. Szlufarska, University of Wisconsin-Madison, Madison, WI, USA}
	
\maketitle

\begin{abstract}
  As beam power continues to increase in next-generation accelerator facilities, high-power target systems face crucial challenges. Components like beam windows and particle-production targets must endure significantly higher levels of particle fluence. The primary beam’s energy deposition causes rapid heating (thermal shock) and induces microstructural changes (radiation damage) within the target material. These effects ultimately deteriorate the components’ properties and lifespan. With conventional materials already stretched to their limits, we are exploring novel materials including High-Entropy Alloys and Electrospun Nanofibers that offer a fresh approach to enhancing tolerance against thermal shock and radiation damage. Following an introduction to the challenges facing high-power target systems, we will give an overview of the promising advancements we have made so far in customizing the compositions and microstructures of these pioneering materials. Our focus is on optimizing their in-beam thermomechanical and physics performance. Additionally, we will outline our ongoing plans for in-beam irradiation experiments and advanced material characterizations. The primary goal of this research is to push the frontiers of target materials, thereby enabling future multi-MW facilities that will benefit various programs in high-energy physics and beyond.
\end{abstract}

\section{INTRODUCTION}

Accelerator target facilities consist of critical high-power target devices such as particle production targets and beam windows, that essentially determine the ability to convert protons into secondary particles to enable High Energy Physics (HEP) experiments. The continuous bombardment of these devices by high-energy high-intensity pulsed beams, makes them challenging to operate and maintain \cite{hurh2011, hurh2013}. In particular, beam-induced thermal shock and radiation damage have been identified as the leading cross-cutting challenges facing high-power target facilities \cite{hurh2012, PASI}, and R$\&$D projects to address these material challenges are underway \cite{RaDIATE, hurh2016}.

Beam-induced thermal shock phenomena arise due to localized energy deposition in the material, caused by very short pulsed-beams (1-10 $\mu$s). The rapidly heated core volume of the material, constrained by surrounding cooler target material, generates a sudden localized region of compressive stress that propagates as stress waves at sonic velocities. These cyclic dynamic stresses can be large enough to cause plastic deformation, fatigue and in some cases fracture of the material. On the other hand, radiation damage disrupts the lattice structure of the material through the displacements of atoms, transmutation, and gas production \cite{Kiselev2016}. Over time, lattice disruptions and defects degrade the bulk material properties and target component health. At Fermilab, the failures of the NT-02 neutrino graphite target and a beryllium primary beam window were attributed to radiation damage effects (embrittlement and swelling) and thermal shock \cite{ammigan2016, bidhar2020, kuksenko2018}.

For similar reasons, various accelerator facilities have had to run at reduced beam powers for extended periods and only achieved their maximum powers on target through R$\&$D of their targets and windows \cite{Hylen2017, hasegawa2017, Kramer2014}. As beam powers continue to increase in next-generation accelerator target facilities, it is important to advance the state of the art in targetry materials to enable future HEP experiments, such as the 2.4-MW LNBF/DUNE and proposed Muon collider \cite{ammigan2022}. In particular, High-Entropy Alloys and Nanofibers are novel classes of materials that offer a new design pathway for improving radiation damage and thermal shock resistance.

\section{HIGH-ENTROPY ALLOYS}

Unlike conventional alloys, High-Entropy Alloys (HEAs) consist of multiple principal constituent elements \cite{Miracle2017, Senkov2018}. The compositionally complex matrix of HEAs provides the opportunity to explore a broader range of novel alloy systems to improve thermomechanical performance and radiation damage resistance. Current studies have already shown examples of HEAs with promising properties and remarkable radiation damage resistance \cite{Tsai2014, Lu2016, Jin2016, Atwani2019}. For accelerator beam windows, we are exploring light-weight elements to produce low-density and primarily single-phase HEAs. The initial composition we investigated was the equimolar CrMnV alloy, which has been demonstrated to exhibit single-phase body-centered-cubic (BCC) structure \cite{Barron2020}. Subsequently, we included additional elements to tweak the microstructure and properties of the alloy.

\begin{figure}[!htb]
   \centering
   \includegraphics*[width=0.8\columnwidth]{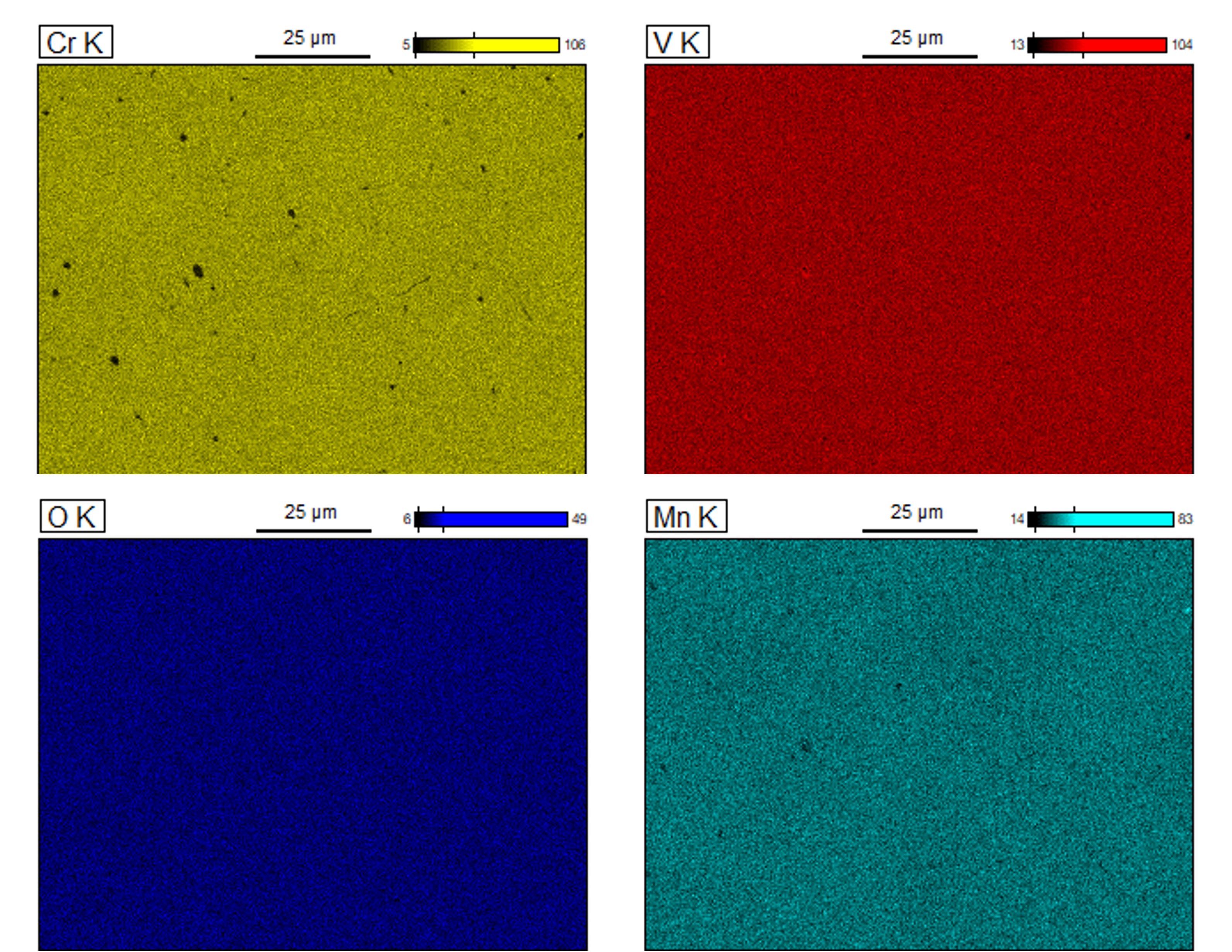}
   \caption{EDS map of CrMnV alloy.}
   \label{fig:EDS}
\end{figure}

The phase diagram of the CrMnV alloy was predicted by CALPHAD \cite{CALPHAD}, before plates were fabricated through the vacuum arc melted method. The as-cast plates underwent hot isostatic pressing (HIP) and homogenization at 1200 $^\circ$C to further promote the single BCC phase. Subsequent characterization utilizing X-Ray diffraction (XRD), scanning electron microscopy (SEM) and energy dispersive spectroscopy (EDS) effectively validated the homogeneity of the alloy, confirming its composition to within 1~$\%$, and the presence of the single-phase BCC structure. Figure \ref{fig:EDS} shows the EDS map of the CrMnV alloy, demonstrating a predominantly homogenized alloy with minimal precipitation.

At the University of Wisconsin Ion Beam Lab, a preliminary irradiation was conducted using heavy ions to replicate displacement damage from high-energy protons. This method offers an efficient and rapid means to evaluate radiation-induced effects and screen materials without inducing activation \cite{Pellemoine2022}. The CrMnV samples were irradiated with 3.7 MeV V$^{2+}$ at 500 $^\circ$C, reaching up to 100 DPA (displacements per atom). Titanium Ti-6Al-4V, commonly used as beam window material, underwent the same irradiation conditions to serve as a benchmark for comparing the performance of the CrMnV HEA.

\begin{figure}[!htb]
   \centering
   \includegraphics*[width=0.8\columnwidth]{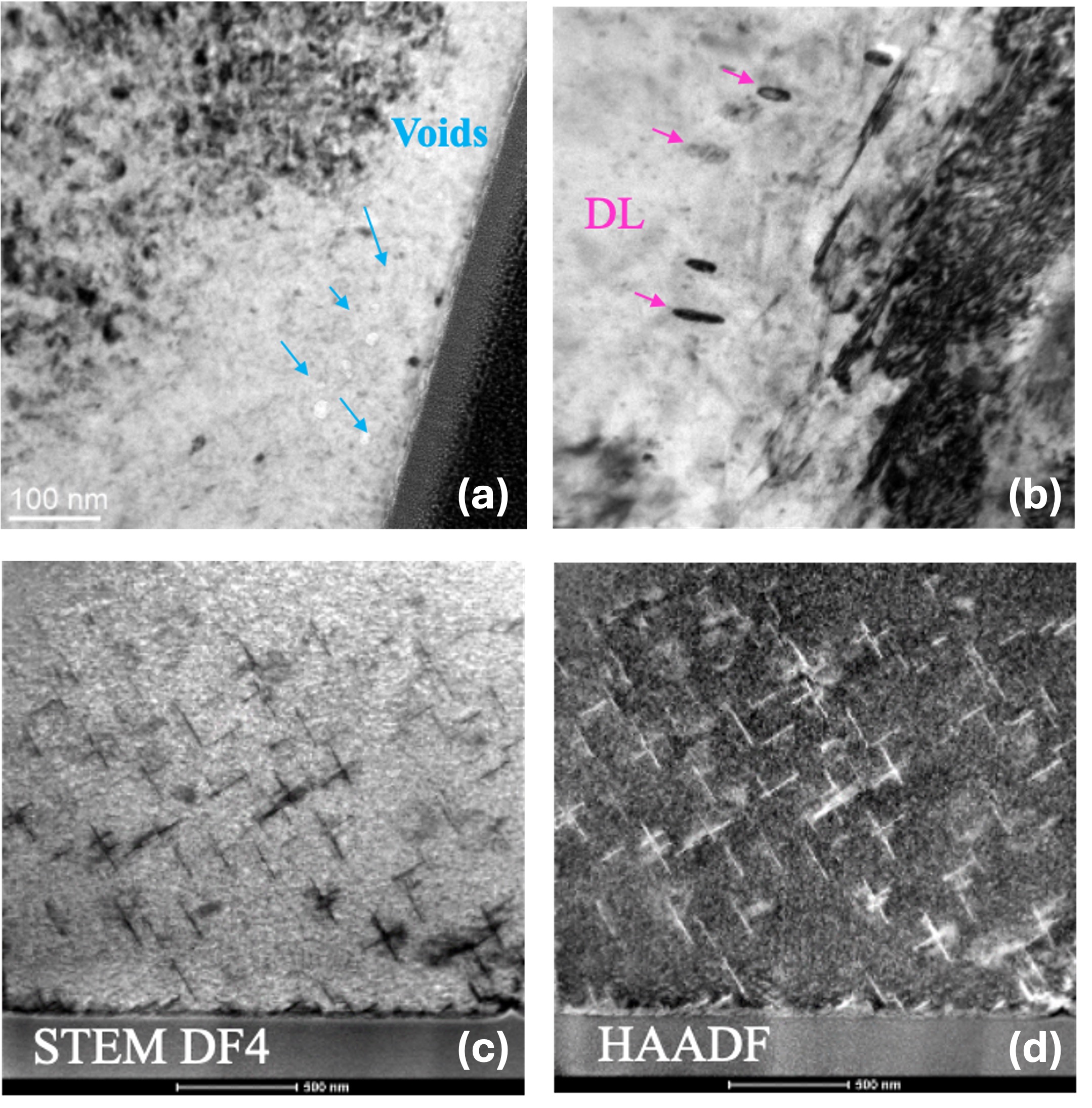}
   \caption{TEM images showing (a,b) voids and dislocation loops (DL) in Ti-6Al-4V, and (c,d) needle morphology of CrMnV HEA without voids or dislocation loops, after 3.7~MeV V$^{2+}$ irradiation at 500~$^\circ$C to 50 DPA.}
   \label{fig:TEM}
\end{figure}

\begin{figure}[!htb]
   \centering
   \includegraphics*[width=0.8\columnwidth]{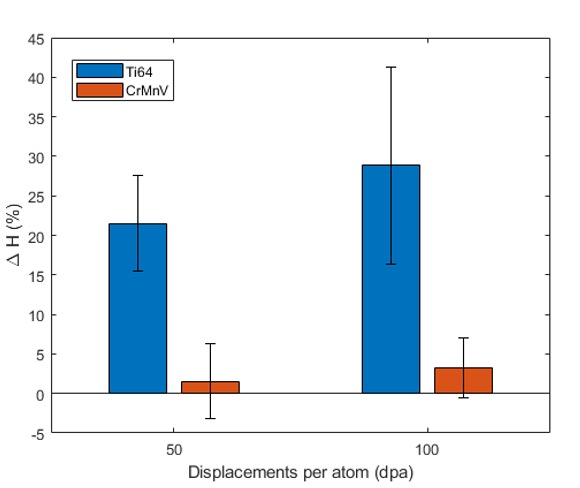}
   \caption{Change in nanoindentation hardness of Ti-6Al-4V and CrMnV HEA upon irradiation up to 100 DPA.}
   \label{fig:Nano}
\end{figure}

Following irradiation, Transmission Electron Microscopy (TEM) analyses were conducted to assess microstructural radiation damage mechanisms, including dislocation loops, voids and radiation-induced segregation. Figure \ref{fig:TEM} shows TEM images of the Ti-6Al-4V and CrMnV alloys irradiated up to 50 DPA. In the Ti-6Al-4V samples, the presence of dislocation loops and voids is clearly observed, whereas the CrMnV HEA exhibits a needle-like morphology without voids or dislocation loops. Scanning TEM EDS analysis revealed no chemical segregation along the needles. Nanoindentation testing was also conducted to assess changes in material hardness upon irradiation. As depicted in \ref{fig:Nano}, minimal hardening was observed in the CrMnV (less than 5$\%$ at 100 DPA) compared to the Ti-6Al-4V (about 30$\%$). The observed formation of voids and dislocation loops is likely contributing to the increased hardening of Ti-6Al-4V in contrast to the CrMnV HEA.

These preliminary findings already suggest enhanced radiation damage resistance in the CrMnV HEA compared to the conventional Ti-6A-4V alloy. Furthermore, the CrMnV HEA was tested with high-energy high-intensity (440 GeV, ~2.4 $\times$ 10$^{15}$ p/cm$^2$ peak fluence) single-shot proton beam pulses at CERN's HiRadMat facility \cite{HRMT} to evaluate its thermal shock resistance. While no visual indication of failure or cracks was observed, planned high-resolution SEM imaging and profilometry will confirm any incurred damage or plastic deformation.

Presently, we are refining the initial CrMnV HEA by adding other elements such as aluminum, titanium, and cobalt, to enhance the microstructure and thermomechanical properties. Our focus lies particularly on stabilizing the BCC phase, reducing impurity concentrations within the solid solution, minimizing precipitate formation, and introducing a semi-coherent B2 phase to enhance alloy strength. To achieve this, we conducted a systematic exploration of compositional space with high-throughput CALPHAD simulations to identify several HEAs (compromising up to 6~components, namely AlCoCrMnTiV) and analyze the effects of the different elements.These alloys have already been synthesized and are currently undergoing microstructural characterization. Furthermore, we intend to measure their bulk physical and mechanical properties.

Multiple heavy-ion irradiation campaigns for the new HEAs are currently in progress. Recently, a low-damage irradiation run (up to 0.4 DPA)  was conducted at the GANIL IRRSUD beamline \cite{IRRSUD} using 36~MeV Ar$^{10+}$ ions. Additionally, preparations are underway for a subsequent irradiation session at higher damage levels (up to 20 DPA) utilizing 4.5~MeV V$^{2+}$ ions at the UW Ion Beam Laboratory. Next year, a thermal shock experiment is planned at the HiRadMat facility, where a subset of HEAs will be exposed to prototypical proton beams. Subsequently, long-term radiation damage studies will be conducted using high-energy prototypical beams for final assessment and selection of novel HEAs for beam windows. Alongside the experimental work, simulations employing Density Functional Theory (DFT) and Molecular Dynamics (MD) are ongoing to elucidate and predict the fundamental behavior of these specific HEAs under irradiation. 

\section{ELECTROSPUN NANOFIBERS}

Electrospun nanofiber material is another novel material we are exploring for secondary-particle production targets, that offers instrinsic resilience to both thermal shock and radiation damage. The nanofiber material continuum, physically discretized at the microscale, enables the fibers to absorb and dampen stress waves. The dimensions of single nanofibers ($\mu$m) that are orders of magnitude smaller than typical beam spot sizes (mm), experience uniform heating (no macroscopic thermal gradients) and thereby reduce thermal shock effects. Discontinuities in the nanofibers further prevent stress waves from propagating throughout the bulk material. Moreover, the nanopolycrystalline structure of nanofibers helps enhance radiation damage resistance, as the associated large number of grain boundaries serves as defect sinks to limit lattice defect growth and mobility. Encouraging outcomes from in-beam tests of ceramic nanofibers at CERN's HiRadMat facility and in-situ ion irradiaiton TEM at the Argonne National Laboratory - IVEM facility have already provided evidence of thermal shock and radiation damage tolerance \cite{Bidhar2021}.

We are now extending the nanofiber technology at Fermilab to impart the desired material properties and physics for secondary production targets, with emphasis on neutrino targets, while also exploring potential synergies within the isotope production community. Our efforts include optimizing the production process of tungsten (W) nanofibers, a high-density material, to compensate for the reduced effective density when in nanofiber form (around 2 g/cm$^3$). The production process involves heating the electrospun W doped polymer nanofiber sample in a furnace with a constant flow of forming gas containing 95$\%$ nitrogen and 5$\%$ hydrogen with a heating rate of 1 $^\circ$C/min. Subsequently, in-situ XRD analysis was conducted at temperatures of 600 $^\circ$C, 750 $^\circ$C and 850 $^\circ$C, with each temperature held constant for 2 hours before analysis.

\begin{figure}[!htb]
   \centering
   \includegraphics*[width=1.0\columnwidth]{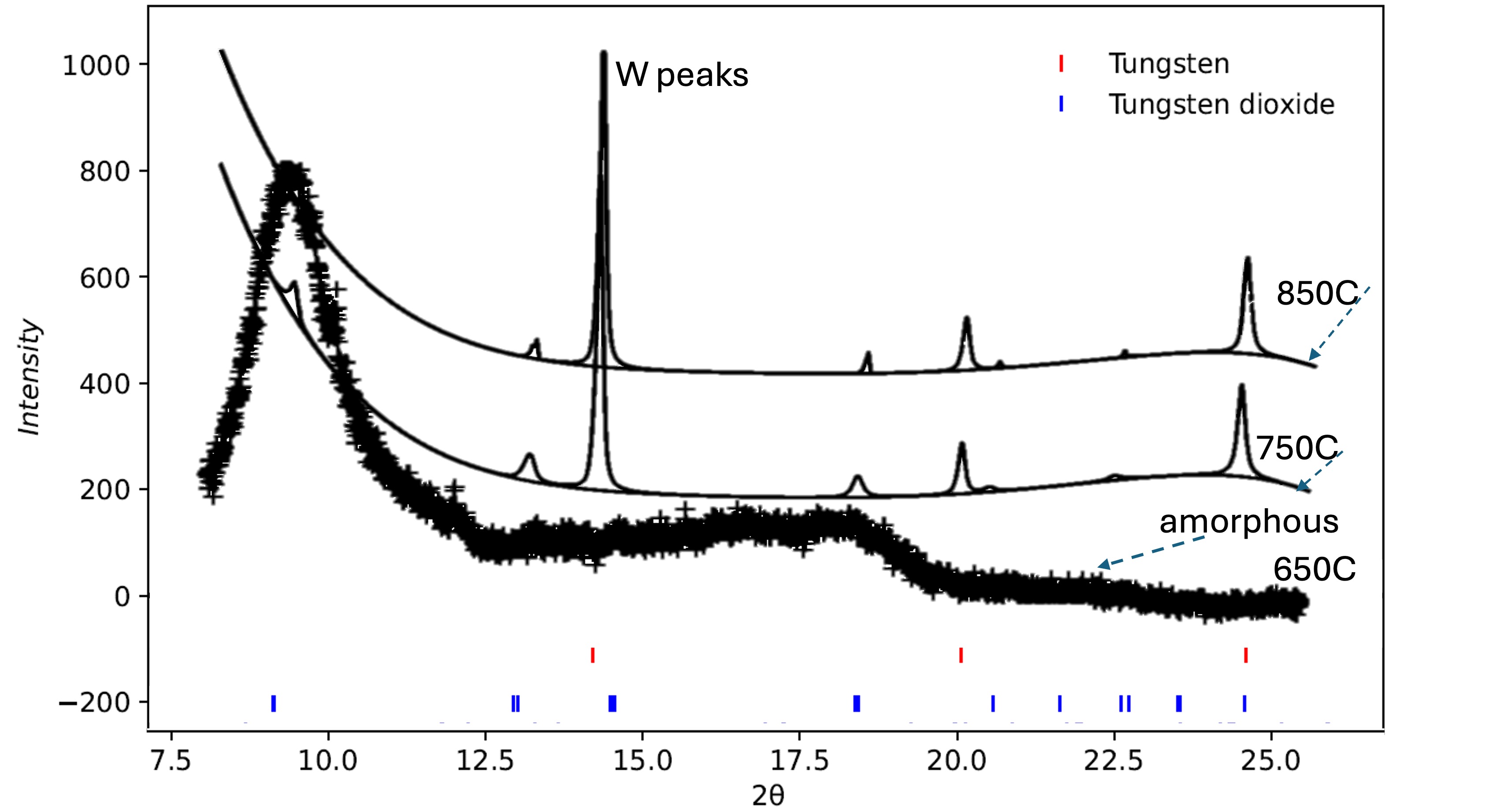}
   \caption{In-situ XRD of tungsten nanofibers during heat treatment.}
   \label{fig:XRD}
\end{figure}

The XRD plot in Figure \ref{fig:XRD} shows the evolution of the W phase. The sample is mostly amorphous at 600 $^\circ$C. As temperature increases to 750$^\circ$C, both W and WO$_2$ (tungsten dioxide) peaks can be seen. Upon increasing the temperature to 850$^\circ$C, the WO$_2$ peaks disappear or fall below the detectable range of the XRD measurement, successfully yielding almost pure W nanofibers (99$\%$W) with average fiber diameter of less than 0.5 $\mu$m. We are currently investigating methods to quantify the degree of crystallinity in the W nanofibers. Concurrently, multiphysics simulations of the nanofibers are underway to understand the impact of changes in the nanofiber packing density on thermal shock response \cite{Asztalos2024}. 


\section{CONCLUSION}
The preliminary efforts and outcomes in the development of high-entropy alloys and nanofiber materials for accelerator target applications demonstrate significant promise. Our overarching objective is to advance the state-of-the art in high-power target materials to enable next-generation multi-MW accelerator target facilities. The novel material development process is iterative in nature and will entail refining the composition, microstructure, and bulk properties. This will be achieved through a series of irradiation experiments followed by comprehensive material characterization, alongside complementary DFT and MD modeling of radiation damage effects.

%
%
\ifboolexpr{bool{jacowbiblatex}}%
	{\printbibliography}%
	{%
	
	
} 

%
%


\end{document}